\documentclass[aps,twocolumn,pra]{revtex4}
\usepackage{amsmath}
\usepackage{amsfonts}
\usepackage{amssymb}
\usepackage{graphicx}
\begin{document}
\title{Harmonic mixing in two coupled qubits: quantum synchronization via ac drives}
\author{S.E. Savel'ev, Z. Washington, A.M. Zagoskin, M.J. Everitt}
\affiliation{$^1$Department of Physics, Loughborough University, Leicestershire, LE11 3TU, United Kingdom}
\begin{abstract}
Simulating  a system of two driven coupled qubits, we show that the time-averaged probability to find one driven qubit in its ground or excited state  can
be controlled by an ac drive in the second qubit. Moreover, off-diagonal elements of the density matrix responsible for quantum coherence can  also be 
controlled via driving the second qubit, i.e., quantum coherence can be enhanced by appropriate choice of the bi-harmonic signal.
Such a dynamic synchronization of two differently driven qubits has an analogy with harmonic mixing  
of Brownian particles forced by two signals through a substrate. Nevertheless, the quantum synchronization in two qubits occurs due to 
multiplicative coupling of signals in the qubits rather than via a nonlinear harmonic mixing for a classical nano-particle. 
\end{abstract}
\maketitle
\section{Introduction}
While the desire of building code-cracking quantum computers \cite{Bellac,Zagoskin2011} remains a long standing goal, its pursuit has pushed forward the enormous progress achieved in quantum mesoscopic physics and quantum nanodevices. These efforts have already resulted in the development of a new class of mesoscopic devices \cite{Nori-new,Nori2,smirnov,mi} and even new types of materials known as quantum metamaterials \cite{metamat,ost-zag}. 

Development of such new devices requires a deep understanding of dynamics of single and multi-qubit systems driven both by ac signals (e.g., electromagnetic radiation, or ac voltage and/or currents) and noise. Following this direction of research, several quantum amplifiers have been recently proposed for one \cite{omelya1} and two-qubit \cite{omelya2,omelya3} systems. The idea \cite{omelya1,omelya2} was to extend the stochastic resonance phenomenon (amplification
by certain amount of noise) for one or two ac driven qubits to enhance quantum coherence. Further analogy \cite{omelya3,rakh,param} between two-qubit system, a Brownian particle driven in periodic substrate and a parametric amplifier has resulted in a proposal to use two coupled qubits as an amplifier
of quantum oscillations.
 
This work was motivated by the analogy between driven Brownian nanoparticles and a system of coupled qubits. It is well-known that an overdamped particle driven by two harmonic ac signals through the substrate can drift in any desirable direction if frequencies
of the two drives are commensurate. This effect known as harmonic mixing \cite{mix1,mix2,mix3,mix4,mix5,mix6,mix7,mix8,mix9} has been already observed in many systems, including
vortices in superconductors \cite{marat}, nanoparticles driven through a pore \cite{nanopore}, current driven Josephson junctions \cite{ustinov} etc. This suggests an idea that a coupled two-qubit system also should exhibit a harmonic mixing behaviour, but in contrast to the usual classical harmonic mixing for overdamped particles, quantum harmonic mixing should be
via parametric coupling of two drives in the quantum master equation. This effect can be used to synchronize quantum oscillations in the two qubits and can control the 
average probability for a qubit to stay in either ground or excited state by changing the relative phase and/or frequencies of the bi-harmonic drive. 

Further, our results can be applied to the case when one needs to control qubits, which do not have their own control circuitry for the reasons of limiting the decoherence brought in by such extra elements, or because of accessibility (e.g., in case of 2D or 3D qubit arrays \cite{everitt,2djosephson} with control circuitry placed at the boundary viz. surface of the device). Similar problems arise in the so-called indirect quantum tomography (see e.g., Ref.\onlinecite{fn1}) or in quantum computations with access to a limited number of qubits (see e.g., Ref. \onlinecite{fn2}). In all these cases, the proposed method of harmonic mixing in qubits allows to control the state of the second (not directly accessible qubit) by varying the frequency and/or phase of the first (accessible) qubit.

\section{Model}

In order to describe a two-qubit system we will use a Hamiltonian in a spin-representation for each qubit with the so called $\sigma_x-\sigma_x$ coupling:
\begin{equation}
H = -\frac{1}{2} \sum_{j=1,2} \left[ \Delta_j\; \sigma^j_z + \epsilon_j(t)\;\sigma^j_x \right] + g\; \sigma^1_x\: \sigma^2_x
\label{eq-ham}
\end{equation}
where $\sigma^j_z$ and $\sigma^j_x$ are Pauli matrices
corresponding to either the first ($j=1$) or the second ($j=2$)
qubit. The tunnelling
splitting energies $\Delta_j$ are usually determined by the geometry
and fabrication details of the specific device, while the
bias energies $\epsilon_j$ can be driven externally.
For simplicity, we consider two identical qubits; that is, we
assume $\Delta_1=\Delta_2=\Delta$.
Let us drive the qubits by a controlled bi-harmonic drive:
\begin{eqnarray}
\epsilon_1(t)=A_{1}\sin(\omega_1 t)\nonumber\\
\epsilon_1(t)=A_{2}\sin(\omega_2 t+\phi)
\label{drive1}
\end{eqnarray}
In other words, each qubit is driven by its own signal and amplitudes, frequencies and relative phase of these two signal
can be varied at will. The question arises if and under what conditions the second qubit can influence the coherence and occupation of the ground and excited states
of the first one. Therefore, we are interested if the second qubit can be used to control the state of the first qubit via dynamic 
synchronization of their quantum oscillations. 

\begin{figure}[btp]
\begin{center}
\includegraphics*[width=8.0cm]{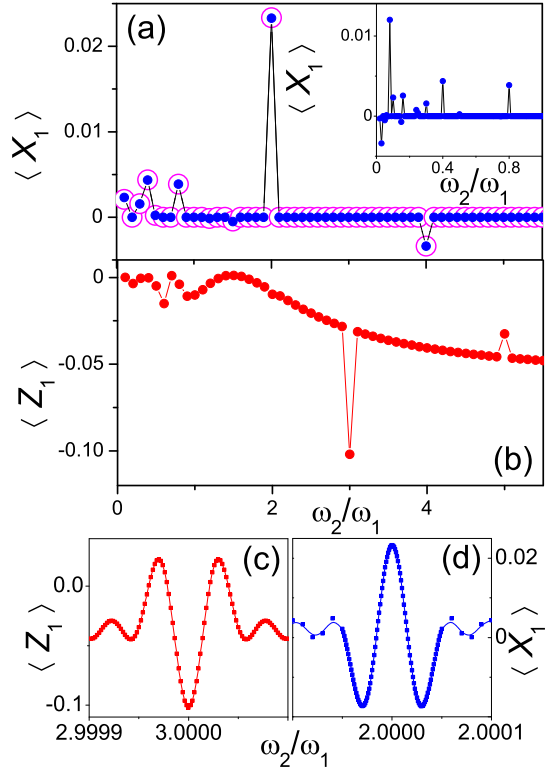}
\end{center}
\caption{Time-averaged Bloch tensor components $\langle X_1\rangle=\langle \Pi_{0x}\rangle$ (a) and $\langle Z_1\rangle=\langle \Pi_{0z}\rangle$ (b) for two coupled qubits driven by the two harmonic 
drives (\ref{drive1}) with parameters: $A_1=A_2=10$, $\phi=0$, $\omega_1=2\sqrt{\Delta^2+g^2}$ and the averaging time interval $5.6\times 10^4<\omega_tt<1.4\times 10^5$ (thus, averaging time $T$ was $8.4\times 10^4/\omega_1$). Other parameters are 
the simulation step $dt=1.13\times 10^{-5}$, the number of simulation steps $5\times 10^9$, damping $\Gamma =10^{-3}$, coupling constant $g=1$, the tunneling
splitting energies $\Delta=1$. To veryfy our numerical results we simulate by using both Euler (open circles) and second-order multidirivative methods (filled circles). 
The time-averaged Bloch tensor element $\langle X_1\rangle$ responsible for the qubit coherence shows peaks at
$\omega_2/\omega_1=2/5, 4/5, 2, 4$, while the time averaged component $\langle Z_1\rangle$ peaks at $\omega_2/\omega_1=3/5, 9/10, 3, 5$. Also, pumping of
the excited state for incommensurate frequencies is clearly seen: $|\langle Z_1\rangle|$ increases for $\omega_2\gtrsim 2\omega_1$. 
By simulating the ten times denser point mesh for frequency ratios from 0.01 to 1 (see inset in (a)) we have obtained extra commensurate resonance frequency ratios but still could not resolve width resonances, which is consistent (see, e.g., 
Ref.~\onlinecite{mix8}) with zero-width harmonic mixing resonances in classical nonlinear devices, where the response near resonances behave as $\cos \Delta\omega T$, where  $T$ is the observation time, and $\Delta\Omega$ is detuning.
In order to check whether we have a similar behaviour for qubit harmonic mixing, we have simulated $\langle Z_1\rangle$ and $\langle X_1\rangle$ (panels (c) and (d)) near resonances $\omega_2/\omega_1=3$ and 2 with frequency detuning $\Delta\omega<1/T$ 
and observed finite resonance width and oscillations near resonances similar to classical harmonic mixing. } \label{fig1}
\end{figure}

The two-qubit density matrix $\hat{\rho}$ can be written as
\begin{equation}
\hat{\rho}=\frac{1}{4} \sum_{a,b=0,x,y,z} \Pi_{ab} \: \sigma^1_a \otimes \sigma^2_b. \label{master}
\end{equation}
This is a straightforward generalization of the standard representation of the single-qubit density matrix expression using the Bloch vector; the components $\Pi_{ab}$ thus constitute what can be called the Bloch tensor. Then, the master equation,
\begin{equation}
\frac{d\hat{\rho}}{dt} = -i\left[\hat{H}(t),\hat{\rho}\right]+\hat{\Gamma}\hat{\rho},
\label{poxyj}
\end{equation}
can be written down directly,
using the standard approximation for the dissipation operator $\hat{\Gamma}$ via the
dephasing $(\Gamma_{\phi 1}, \Gamma_{\phi2})$, and relaxation $(\Gamma_{1}, \Gamma_{2})$
rates, to characterize the intrinsic noise in the system. The master equation (\ref{master}) can be explicitly written as follows \cite{omelya2,omelya3}:
\begin{eqnarray}
\begin{array}{lll}
\dot{\Pi}_{0x} & = & \Delta_2\Pi_{0y} - \Gamma_{\phi 2}\Pi_{0x} \\
\dot{\Pi}_{0y} & = & -\Delta_2\Pi_{0x} + \epsilon_2(t)\Pi_{0z} - 2g\Pi_{xz} - \Gamma_{\phi 2}\Pi_{0y}\\
\dot{\Pi}_{0z} & = & -\epsilon_2(t)\Pi_{0y} + 2g\Pi_{xy}- \Gamma_{2}(\Pi_{0z}-Z_{T2})\\
& & \\
\dot{\Pi}_{x0} & = & \Delta_1\Pi_{y0} - \Gamma_{\phi 1}\Pi_{x0} \\
\dot{\Pi}_{y0} & = & -\Delta_1\Pi_{x0} + \epsilon_1(t)\Pi_{z0} - 2g\Pi_{zx} - \Gamma_{\phi 1}\Pi_{y0}\\
\dot{\Pi}_{z0} & = & -\epsilon_1(t)\Pi_{y0} + 2g\Pi_{yx}- \Gamma_{1}(\Pi_{z0}-Z_{T1})\\
& & \\
\dot{\Pi}_{xx} & = & \Delta_2\Pi_{xy} + \Delta_1\Pi_{yx} - (\Gamma_{\phi 1} + \Gamma_{\phi 2})\Pi_{xx} \\
\dot{\Pi}_{xy} & = & -2g\Pi_{0z} -\Delta_2\Pi_{xx} + \Delta_1\Pi_{yy} \\
&+& \epsilon_2(t)\Pi_{xz} - (\Gamma_{\phi 1} + \Gamma_{\phi 2})\Pi_{xy}\\
\dot{\Pi}_{yx} & = & -2g\Pi_{z0} -\Delta_1\Pi_{xx} + \Delta_2\Pi_{yy} \\
&+& \epsilon_1(t)\Pi_{xz} - (\Gamma_{\phi 1} + \Gamma_{\phi 2})\Pi_{yx}\\
\dot{\Pi}_{xz} & = & 2g\Pi_{0y} - \epsilon_2(t)\Pi_{xy} + \Delta_1\Pi_{yz} - (\Gamma_{\phi 1}+\Gamma_{2})\Pi_{xz}\\
\dot{\Pi}_{zx} & = & 2g\Pi_{y0} - \epsilon_1(t)\Pi_{yx} + \Delta_2\Pi_{zy} - (\Gamma_{\phi 2}+\Gamma_{1})\Pi_{zx}\\
\dot{\Pi}_{yy} & = & -\Delta_1\Pi_{xy} - \Delta_2\Pi_{yx} \\
&+& \epsilon_2(t)\Pi_{yz} + \epsilon_1(t)\Pi_{zy} - (\Gamma_{\phi 1} + \Gamma_{\phi 2})\Pi_{yy}\\
\dot{\Pi}_{yz} & = & - \Delta_1\Pi_{xz} - \epsilon_2(t)\Pi_{yy} \\
&+& \epsilon_1(t)\Pi_{zz} - (\Gamma_{\phi 1}+\Gamma_{2})\Pi_{yz}\\
\dot{\Pi}_{zy} & = & - \Delta_2\Pi_{zx} - \epsilon_1(t)\Pi_{yy} \\
&+& \epsilon_2(t)\Pi_{zz} - (\Gamma_{1}+\Gamma_{\phi 2})\Pi_{zy}\\
\dot{\Pi}_{zz} & = & -\epsilon_1(t)\Pi_{yz} -\epsilon_2(t)\Pi_{zy} \\
&-& (\Gamma_{1} + \Gamma_{2})(\Pi_{zz}-Z_{T1}Z_{T2})
\end{array}
\label{poxy}
\end{eqnarray}
Also, for simplicity,
hereafter we assume that the relaxation rates are the same for both identical
qubits, i.e., $\Gamma_{\phi 1} = \Gamma_{\phi 2}=\Gamma_{\phi}$ and
$ \Gamma_{1}=\Gamma_{2}=\Gamma_{r}$, and the temperature is low enough,
resulting in $ Z_{T2}= Z_{T1}=1$, where $Z_{Tj} = \tanh(\Delta_j/2k_BT_j)$
is the equilibrium value of the $z$-component of the Bloch vector. This set of ordinary differential equations is an ideal starting point 
for numerical analysis of the dynamics of two driven and dissipative qubits as it has been proved before \cite{omelya2,omelya3}. We will simulate these
differential equations for two differently driven qubits and will study quantum harmonic mixing. 

In the limit of zero coupling, $g = 0$, there exists a solution of
Eqs.~(\ref{poxy}) with no entanglement between the qubits. This
solution can be written as a direct product of two independent
density matrices expressed through their Bloch vectors:
\begin{equation}
\hat{\rho}_j = \frac{1}{2} (1+X_j\sigma_x+Y_j\sigma_y+Z_j\sigma_z).
\end{equation}
 The components of the Bloch tensor $\Pi_{ab}$ are all zero with the exception of
 \begin{equation}
(\Pi_{ox},\Pi_{oy},\Pi_{oz})= (X_1,Y_1,Z_1); \:\: (\Pi_{xo},\Pi_{yo},\Pi_{zo})=(X_2,Y_2,Z_2)\; ,
\label{eq:qomponents}
\end{equation}
which are just the separate qubits' Bloch vector components. The density matrix components $\Pi_{ox}=X_1$, $\Pi_{xo}=X_2$, 
$\Pi_{oy}=Y_1$, $\Pi_{yo}=Y_2$, $\Pi_{oz}=Z_1$, and $\Pi_{zo}=Z_2$ can be often directly accessible in experiments. For instance, 
for two coupled flux qubits the circulating currents in each qubit are proportional to $X_1$ or $X_2$, respectively (see e.g., Ref. \onlinecite{omelya1}), 
while $Z_1$ and $Z_2$ determine the occupation probabilities of the upper (lower) level for the first and the second qubit:
\begin{equation}
P_{\pm,j}=\frac{1}{2}\left(1\mp Z_j(t)\right) \label{prob}
\end{equation} 
with $j=1$ or 2.

\section{Simulation results}

We simulated the set (\ref{poxy}) by using the standard Euler method which has been proved to converge well for low-noise drives \cite{omelya2,omelya3} and analyzed 
the time-averaged diagonal element of density matrix $\langle X_1\rangle=\langle\Pi_{ox}\rangle=\lim_{T\rightarrow\infty}\int_0^T\Pi_{ox}dt/T$, responsible for the mean coherence in the first qubit, as well as the time-averaged density matrix element $\langle Z_1\rangle=\langle\Pi_{oz}\rangle=\lim_{T\rightarrow\infty}\int_0^T\Pi_{oz}dt/T$, responsible for the mean occupation 
of the ground and excited states in the first qubit. To verify validity of our numerical results we have also used  higher-order multiderivative
methods to prove the stability of our numerical procedure (compare the open circles for Euler methods and the filled circles for the second order method in Fig. 1 and 2).

\begin{figure}[btp]
\begin{center}
\includegraphics*[width=8.0cm]{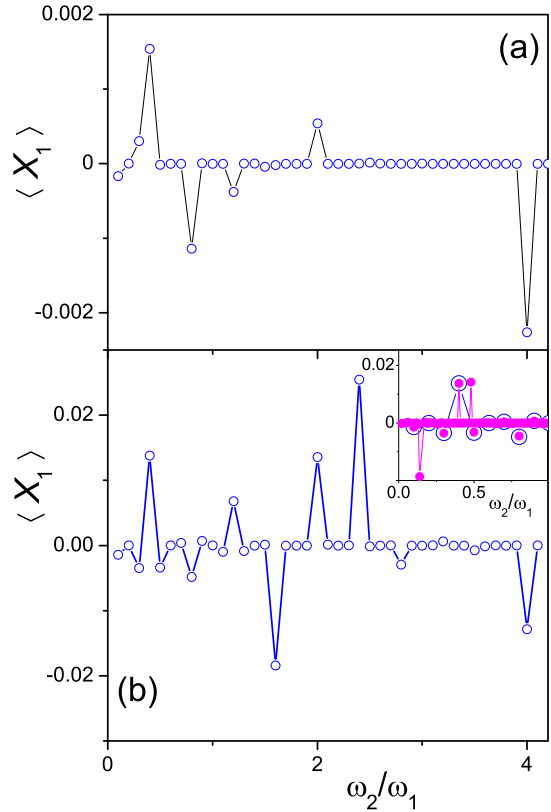}
\end{center}
\caption{Time-averaged Bloch tensor components $\langle X_1\rangle=\langle \Pi_{0x}\rangle$ for two coupled qubits driven by the two harmonic 
drives (Eq.(\ref{drive1})) with the same parameters as in Fig. 1 and driving frequency $\omega_1=\sqrt{\Delta^2+g^2}-g$ (a) equal to 
an energy level transition frequency \cite{omelya3} and for $\omega_1=2.113(\sqrt{\Delta^2+g^2}-g)$ (b) which is away from the energy level transition. 
Simulations with a ten-time denser point mesh for frequency ratio from 0.01 to 1 (see inset in (a)) uncovered
some extra commensurate frequency ratio where peaks in $\langle X_1\rangle$ occurs. }\label{fig2}
\end{figure}
\begin{figure}[btp]
\begin{center}
\includegraphics*[width=8.0cm]{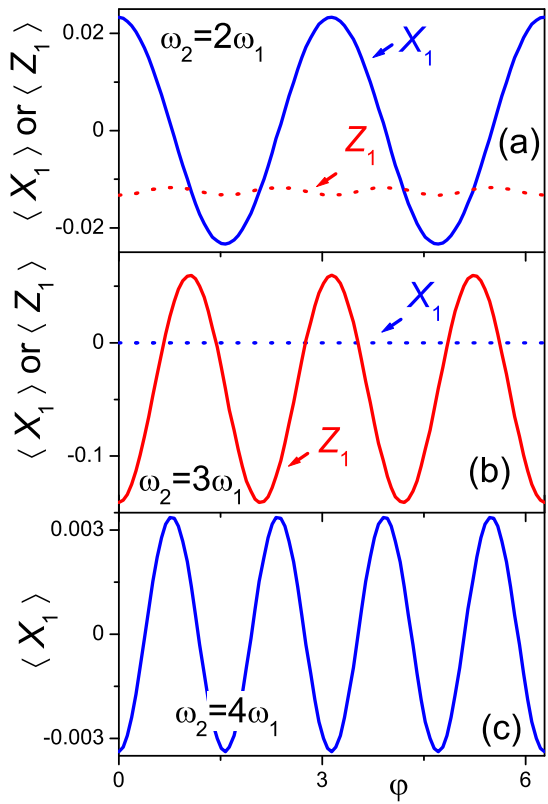}
\end{center}
\caption{Dependence of $\langle X_1\rangle$ and $\langle Z_1\rangle$ on the relative phase of two drives of the bi-harmonic signal at different
frequency ratio $\omega_2/\omega_1=2$ (a), 3(b), 4(c). All other parameters are the same as in Fig. 1. For even frequency ratio,
where $\langle X_1\rangle$ has a peak (Fig.1), the strong dependence of $\langle X_1\rangle(\varphi)$ and a week dependence of $\langle Z_1\rangle(\varphi)$
occurs, while, for odd ratio of $\omega_2/\omega_1$,  dependence of $\langle Z_1\rangle(\varphi)$ is clearly seen and $\langle X_1\rangle(\varphi)$
is negligible. The periods of functions $\langle X_1\rangle(\varphi)$ and $\langle Z_1\rangle(\varphi)$ are controlled by the frequency ratio and are equal to
$2\pi\omega_1/\omega_2$ (for $\omega_2>\omega_1$).  }\label{fig3}
\end{figure}

As we expected, there is no mean coherence $\langle X_1\rangle\approx 0$ for most of frequency ratio $\omega_1/\omega_2$ 
apart from the specific commensurate cases (e.g., $\omega_1/\omega_2=2/5, 4/5, 2, 4$, see Fig.1a). Such a situation reminds of a usual classical 
harmonic mixing for nanoparticles (see, e.g., \cite{mix8}), however, the frequency ratios where peaks occur, are also tuneable by changing the absolute value of
signal frequency in either the first or the second qubit. Indeed, choosing the frequency $\omega_1$ to be equal to the inter-level spacing frequency $\omega_1= \sqrt{\Delta^2+g^2}- g$ (Fig. 1a) or $\omega_1=2\sqrt{\Delta^2+g^2}$ (Fig.2a) or even away from the inter-level resonances 
$\omega_1=2.113(\sqrt{\Delta^2+g^2}-g)$ (Fig. 2b) results in a qualitatively similar peak structure, but showing different sequence of the frequency ratios. Indeed, for $\omega_1=2.113(\sqrt{\Delta^2+g^2}-g)$ (Fig. 2b),
several new frequency ratios corresponding to the enhancement of qubit coherence occur at $\omega_2/\omega_1=2/5, 4/5, 6/5, 8/5, 2, 12/5, 14/5, 4$. Moreover, some peaks can even change their signs (compare peaks at $\omega_2/\omega_1=4/5$ in Fig. 1a, 2a, 2b) indicating that both the frequency ratio 
and the absolute value of frequency can be used to tune qubit harmonic mixing. 
Such a behaviour is quite unusual with respect to classical harmonic mixing (see e.g., \cite{mix3})
where the frequency ratio is defined by nonlinearity of the system. In contrast, the master equation set (\ref{poxy}) is linear and harmonic mixing occurs via a
mixture of multiplicative drives as in the qubit parametric oscillator. As we have recently shown, this results in a quite unusual spectra of $\Pi_{a,b}$, and in particular $X_1$ and $Z_1$ with many harmonic peaks showing complex hierarchy. This can explain a non-trivial behaviour of harmonic mixing changes when
varying $\omega_1$ or $\omega_2$. Note also, that the quantum harmonic mixing occurs in both cases: when (i) $\omega_1$ is equal to inter-level
spacing and (ii) away from this situation. Therefore, there is no need to tune the parameters of the external drives to any characteristic internal frequency
of the two qubit system to observe quantum harmonic mixing.

We have also observed a similar harmonic mixing in time-averaged matrix element $\langle Z_1\rangle$ 
responsible for the occupation of the excited and ground states (Fig. 1b). Interestingly, the peaks 
in $\langle Z_1\rangle$ occur at different ratios of bi-harmonic drive $\omega_2/\omega_1=3/5,9/10,3, 5$. However, 
such a behaviour is perfectly consistent with the harmonic spectra of $\langle Z_1\rangle$ and $\langle X_1\rangle$ studied in \cite{omelya3}.
Indeed, the specrum of $X_1$ contains only odd harmonics, while the spectrum of $Z_1$ consist of even harmonics in agreement with the fact that peaks of $\langle X_1\rangle$ and $\langle Z_1\rangle$ have a different parity. Moreover, apart from the peaks at the specific commensurate frequencies, the value of $|\langle Z_1\rangle|$
gradually increases with the frequency for $\omega_2\gtrsim 2\omega_1$ indicating pumping in the excited state even for incommensurate frequencies. 

Following the analogy with classical harmonic mixing \cite{mix8}, we expect the dependence of  $\langle X_1\rangle$ and $\langle Z_1\rangle$ on 
the relative phase $\varphi$ of bi-harmonic drive at commensurate 
frequencies where peaks have been observed. Indeed, we obtained such a dependence
$\langle X_1\rangle(\varphi)$ and $\langle Z_1\rangle(\varphi)$ shown in Fig. 3(a-c) for the same simulation parameters as in Fig. 1 and for
frequency ratios $\omega_2/\omega_1=2$ (a), 3 (b), 4 (c). The well resolved peaks of $\langle X_1\rangle$ at even frequency ratios $\omega_2/\omega_1=2$ and 4 exhibit a strong dependence on relative phase, 
while very weak peaks of $\langle Z_1\rangle$ at these frequency ratios show almost no dependence on $\varphi$. Comparing figures 3(a) and 3(c), we also conclude that periodicity of the $\langle X_1\rangle(\varphi)$ changes with increasing frequency ratio following the rule:
$\langle X_1\rangle(\varphi+2\pi \omega_1/\omega_4)=\langle X_1\rangle(\varphi)$. Therefore, the number of full oscillations increases with frequency
ratio $\omega_2/\omega_1$. This dependence of the $\varphi$-periods of $\langle X_1\rangle$ and $\langle Z_1\rangle$ oscillations on the frequency ratio of
the harmonic drives is analogous to the similar dependence of classical harmonic mixing of a Brownian particle driven 
by bi-harmonic drive on nonlinear substrate \cite{mix8}.

\section{Conclusions}

We have predicted quantum harmonic mixing in a two-qubit system driven by a  bi-harmonic drive. It manifests itself in a set of peaks of time-averaged
density matrix components responsible for both qubit coherence and occupation of ground and excited states
of the qubits. These peaks can be controlled not only by
the ratio of frequencies of the two signals but also by tuning frequencies themselves and by relative phase of the two signals. 
Such a quantum harmonic mixing can be used
to manipulate one driven qubit by applying an additional ac signal to the other qubit coupled with the one we have to control.
Indeed, setting the frequency of the second qubit to be three times larger than the one of the first qubit and changing the relative phase of signals in these qubits
produces oscillations of the density matrix element $Z_1$ with amplitude of about 0.1 according to Fig. 3b. Therefore, by changing the driving signal only 
in the second qubit should allow us to indirectly vary the occupation probabilities of the upper level in the first qubit between 47 and 58 percent (see eq. (\ref{prob})). A stronger
coupling between the qubits should allow an even larger amplitude of the indirect control of the occupation probability. This effect is obviously robust to a 
reasonably strong decoherence and energy dissipation in the system. 
\vspace{-0.5cm}
\acknowledgements
\vspace{-0.5cm}
S.S., A.Z., and M.E. acknowledge that this publication
was made possible through the support of a grant
from the John Templeton Foundation; the opinions expressed
in this publication are those of the authors and
do not necessarily reflect the views of the John Templeton
Foundation. S.S. also acknowledges The Leverhulme 
Trust for partial support of this research.

\end{document}